\documentclass[journal]{IEEEtran}
    
\usepackage{amsmath}
\usepackage{todonotes}
\usepackage{colortbl}
\usepackage{hhline}
\usepackage{subfigure}
\usepackage{graphicx}
\usepackage{threeparttable,tabularx}
\usepackage{booktabs}
\usepackage{multirow}
\usepackage{url}                
\usepackage{array}
\usepackage{tcolorbox}
\usepackage{ragged2e}
\usepackage{pifont}
\usepackage{algorithm}
\usepackage{algpseudocode}
\usepackage{hyperref}

\begin{document}

\definecolor{SoftRed}{rgb}{0.9, 0.6, 0.6} 
\definecolor{SoftAmber}{rgb}{1.0, 0.9, 0.7} 
\definecolor{SoftGreen}{rgb}{0.7, 0.9, 0.7} 
\definecolor{PaleBlue}{rgb}{0.8, 0.9, 1.0} 

\newcommand{\cellcolorme}[1]{%
  \ifnum #1>89 \cellcolor{SoftRed}\fi
  \ifnum #1<90 \ifnum #1>79 \cellcolor{SoftAmber}\fi\fi
  \ifnum #1<80 \ifnum #1>59 \cellcolor{SoftGreen}\fi\fi
  \ifnum #1<60 \cellcolor{PaleBlue}\fi
  #1
}
\title{Attacking Delay-based PUFs with Minimal Adversarial Knowledge} 

\author{Hongming Fei, Owen Millwood, Prosanta Gope \textit{Senior Member, IEEE}, Jack Miskelly, Biplab Sikdar \textit{Senior Member, IEEE}%
 \thanks{H. Fei and B. Sikdar are with the Department of Electrical and Computer Engineering, National University of Singapore, Singapore.}%
 \thanks{P. Gope and O. Millwood are with the Department of Computer Science, The University of Sheffield, Regent Court, Sheffield S1 4DP, United Kingdom.}
\thanks{J. Miskelly is with the Centre for Secure Information Technologies, Queen's University Belfast, United Kingdom.}
\thanks{(E-mail: fei.hongming@u.nus.edu, p.gope@sheffield.ac.uk)}
}

\markboth{IEEE Transactions on Information Forensics and Security}%
{Shell \MakeLowercase{\textit{et al.}}: Attacking PUFs with Minimal Adversarial Knowledge}
\maketitle
\begin{abstract}
 Physically Unclonable Functions (PUFs) provide a streamlined solution for lightweight device authentication. Delay-based Arbiter PUFs, with their ease of implementation and vast challenge space, have received significant attention; however, they are not immune to modelling attacks that exploit correlations between their inputs and outputs. Research is therefore polarized between developing modelling-resistant PUFs and devising machine learning attacks against them. This dichotomy often results in exaggerated concerns and overconfidence in PUF security, primarily because there \textbf{lacks} a universal tool to gauge a PUF's security. In many scenarios, attacks require additional information, such as PUF type or configuration parameters. Alarmingly, new PUFs are often branded `secure' if they lack a specific attack model upon introduction. To impartially assess the security of delay-based PUFs, we present a \textbf{generic} framework featuring a Mixture-of-PUF-Experts (MoPE) structure for mounting attacks on various PUFs with \textbf{minimal adversarial knowledge}, which provides a way to compare their performance fairly and impartially. We demonstrate the capability of our model to attack different PUF types, including the \textit{first} successful attack on \textbf{Heterogeneous Feed-Forward PUFs} using only a reasonable amount of challenges and responses. We propose an extension version of our model, a Multi-gate Mixture-of-PUF-Experts (MMoPE) structure, facilitating multi-task learning across diverse PUFs to recognise commonalities across PUF designs. This allows a streamlining of training periods for attacking multiple PUFs simultaneously. We conclude by showcasing the potent performance of MoPE and MMoPE across a spectrum of PUF types, employing simulated, real-world unbiased, and biased data sets for analysis.
\end{abstract}


\IEEEpeerreviewmaketitle

\section{Introduction}
\label{sec:intro}
Lightweight authentication is gaining momentum as a vital research area, primarily due to the ubiquitous integration of the Internet of Things (IoT) in our daily lives. Numerous methodologies have emerged to bolster its resilience. Traditionally, secret keys stored in non-volatile memory are employed to encrypt sensitive data, and cryptographic techniques, such as asymmetric cryptography, have been used for device authentication \cite{Reference1}. However, cryptographic implementations can be resource-intensive, particularly given the nature of IoT devices, which are often resource-constrained. Even with cryptography, devices remain susceptible to various threats, including invasive attacks \cite{Tria2011}. In contrast, Physical Unclonable Functions (PUFs) offer a streamlined approach to security, suitable for both authentication and secure key generation. Their efficacy in authentication revolves around optimal energy consumption, computational power, and robust defence against threats. This is largely attributed to PUFs deriving volatile secrets from a device's inherent physical characteristics rather than relying on stored secrets in non-volatile memory. These characteristics, resulting from random variations during integrated circuit (IC) manufacturing, ensure that no two ICs are identical \cite{Reference1}. Moreover, the unique delay sequences in transistors and wires of each IC make PUFs capable of generating unpredictable sequences, offering a formidable defence against malicious attacks. Furthermore, PUFs are efficient, negating the need for intricate cryptographic operations.\\

PUFs utilize sequences of binary numbers as input and output, referred to as challenge-response pairs (CRPs) \cite{Reference2}. Based on the number of CRPs they can produce, PUFs are classified as `Weak' and `Strong'. This terminology has no bearing on the security properties of the PUF but rather indicates the total number of supported unique CRPs. While Weak PUFs are limited to generating a linear number of CRPs, making them apt for key generation and storage, Strong PUFs can produce an exponentially growing number of unique CRPs (based on PUF size), making them ideal for creating one-time authentication tokens. However, a key vulnerability persists for PUFs, particularly for delay-based Arbiter PUFs, known as `machine learning modelling attacks' (ML-MA). In such attacks, adversaries collect CRPs produced by PUFs and employ machine learning techniques to deduce the challenge-response correlation \cite{Ruhrmair_modelling}. As such a model can predict the response to future challenges, it can allow an adversary to pose as the PUF-authenticated device. The high prediction accuracy of this technique jeopardizes the security of PUFs. Consequently, diverse PUF variations have been proposed to fortify their defences. For instance, XOR Arbiter PUFs (XOR-APUFs) integrate the responses of multiple Arbiter PUFs (APUFs) \cite{Reference4}\cite{Reference5}, while (XOR) Feed-Forward Arbiter PUFs (FF-APUFs) incorporate a ``feed-forward loop" (FF-loop) concept and Interpose PUF (iPUF) utilizes the upper-lower-layers design. All are geared towards enhancing PUF non-linearity \cite{Ruhrmair_modelling}. Nevertheless, so long as the training CRP set is large enough, accurate predictions of these complex PUFs are still feasible. Without more extreme countermeasures, they only raise the modelling effort and do not fundamentally prevent it.

\subsection{Related Work}
\label{sec:related}

In general, the crux of modelling attacks on PUFs is identifying the relationship between challenges and responses. This implies that the attacker needs access to the CRPs in the PUF. Multiple studies have explored how to conduct modelling attacks on PUFs using ML methods. Rührmair et al. first employed Logistic Regression (LR) to model XOR-PUFs \cite{Ruhrmair_modelling}. They began by formalizing the mathematical model of the arbiter PUF and then captured the XOR logic for the machine learning model. For instance, they utilized multiplication operations of ${1,-1}$ to symbolize XOR, achieving high modelling accuracy. Yet, their mathematical model was less effective when faced with noisy data and intricate PUFs, such as those with many XOR-APUF stages. In \cite{Ruhrmair_LR_ES}, both LR and Evolution Strategies (ES) were deployed to target various PUFs (some of which are examined in this paper) using both simulated and silicon data. Their outcomes reaffirmed the viability of ML-MA. PUFs' reliability also emerged as a vulnerability, revealing differences in delay paths. Becker \cite{Becker_gap} introduced reliability-based machine learning attacks, demonstrating a link between response reliability and delay differences, suggesting that an unstable response indicated a minuscule delay difference.
In \cite{Reference14}, researchers presented a comprehensive framework to conduct logical approximation modelling attacks on several delay-based PUFs. The framework leveraged a logical approximation and global approximation rooted in artificial neural networks (ANN). The former estimated basic logical operations like AND and OR in circuits, applied linear functions, and built an ANN model of the logical architecture. For more intricate logics, the global approximation discerned an appropriate continuous function to map the challenge-response relation, subsequently formulating an ANN structure to emulate the chosen function. The primary goal of \cite{Reference14} was to depict the nonlinear components in PUF designs through soft models to expedite modelling.
In \cite{Reference4}, the authors delved into another type of modelling attack on Arbiter PUF compositions. This attack was founded on a deep feed-forward neural network and did not rely on the mathematical model or structural knowledge of PUFs as seen in prior work \cite{Ruhrmair_modelling,Ruhrmair_LR_ES,Becker_gap}. Distinct layers with varying neurons were meticulously designed to enable a successful attack. Aseeri et al. \cite{aseeri2018machine} considered a different neural network approach to model XOR PUFs. This structure is more standardized, expanding with the increase of component arbiter PUFs. Mursi et al. in \cite{Mursi} refined the structure and various ML hyper-parameters, outperforming \cite{Reference4,aseeri2018machine}. Nonetheless, the efficiency of ML-MA methods varies across devices and datasets. 

In \cite{wisiol2022neural}, Wisiol et al. provided unbiased comparisons of recent ML-MAs, underscoring the promise of neural network-based attacks on PUFs. They also unveiled a new model for XOR FF-APUFs but admitted their inability to breach heterogeneous XOR Feed-Forward Arbiter PUFs, even under moderate parameter settings. A recurring observation across these studies was the sensitivity of their methods to model settings. Even minor alterations could result in attack failures, indicating designs tailored for specific PUFs. In \cite{mishra2024calypso}, a new evolutionary search (ES) based modelling method $\mathrm{CalyPSO}$ was proposed. $\mathrm{CalyPSO}$ targets delay-based PUFs and was able to mount successful modelling attacks on high-order XOR-PUFs, interpose-PUFs and LR-PUFs. The method requires knowledge of the architectural topologies, and the training cost is exceptionally high relative to other approaches. Significantly, compared to this work, they attempt a ``downgrade attack" where they attack $n-XOR$ PUFs using a model tuned for $(n+1)-XOR$ and achieve moderate success, making this method at least partially generic. While the concepts in this paper are interesting, we have some reservations about the results claimed due to errors found in the codebase associated with the paper and the resulting difficulty in replicating the results, e.g., overlaps between the training dataset and test dataset, miscalculation of the amount of training data, etc.

Apart from CRPs, side channel information is also used for modelling PUFs. In \cite{liu2022alsca}, an auxiliary learning framework $\mathrm{ALScA}$ is proposed, which is a multiple-task learning model utilizing both CRPs and side channel information to help modelling tasks. $\mathrm{ALScA}$ has a shared-bottom structure and focuses on the correlations between the mathematical model and side-channel models and achieves good accuracy improvement on Arbiter-based PUFs.

\subsection{Problem Statement and Motivation}
\begin{figure}[t!]
\begin{center}
\makebox[0pt]{\includegraphics[width=\linewidth]{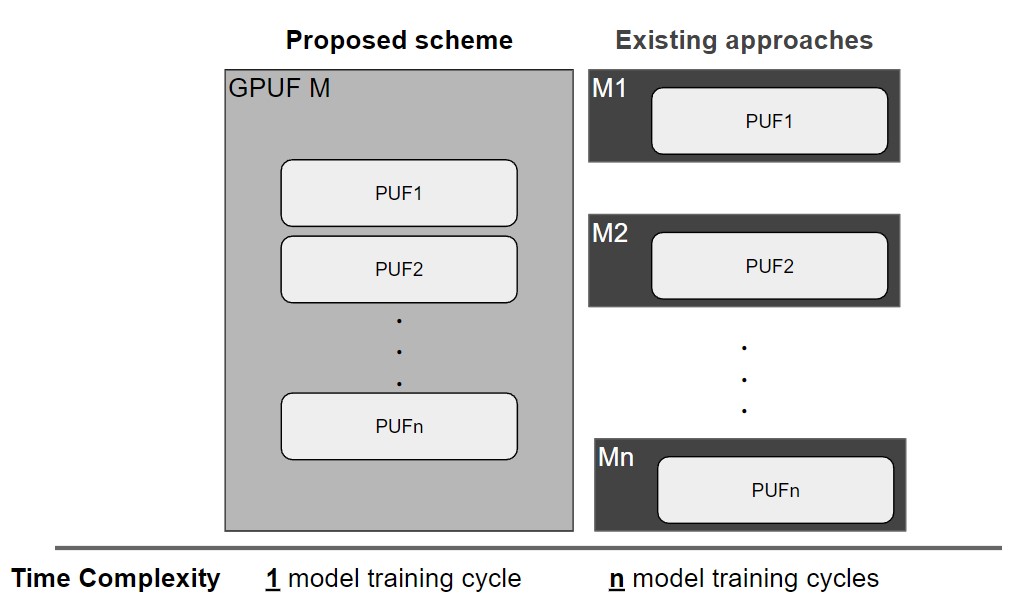}}
\end{center}
\caption{Generic Framework vs. Multiple single models for PUF modelling}
\label{fig:motivation}
\end{figure}


Strong PUF-based authentication threat models often assume an attacker capable of gathering large numbers of CRPs and a single target PUF of fixed type. Further, the structure of the PUF is known to the adversary. In many ways, this is a worst-case scenario for remote attacks: a logical threat model for most PUF papers where the aim is either to prove the security of a given design or to disprove such claims previously made. In reality, it's unlikely that most attackers will have unlimited ability to read network traffic or such detailed knowledge of the hardware implementation. If, indeed, the case is be only one PUF exists in the network. PUF authentication protocols are generally agnostic of the specific hardware implementation. So, there is no reason to assume that every device has the same PUF or PUFs in a mixed device network. This isn't something that can be easily inferred from network traffic alone, nor is it something that manufacturers are likely to disclose. Thus, when attacking PUFs, we should treat them as the real black box; only input and output data are available and no other information. Besides, regardless of whether a PUF is provably secure, in relative terms, what is the real risk from an opportunistic attacker? The kind of attacker simply probing for weaknesses, who doesn't have unlimited ability to avoid detection, who isn't a hardware security expert but has enough technical skill to use any tools that such experts have produced. Does such an adversary pose a real threat with the tools currently available? If not, could a tool be built which enables that threat? We should expect the adversary to prefer to attack without the need to know much more details about PUFs other than CRPs. As discussed above, this kind of extra information can be hard to obtain in practical scenarios.


Prevailing ML-MA techniques necessitate information beyond just CRPs. We classify this information into `Explicit' and `Implicit' categories. Explicit information encompasses setup parameters of PUFs, such as the placement of Feed-Forward Loops. On the other hand, Implicit information refers to type-specific details, like the count of APUFs in XOR PUFs or the number of Feed-Forward Loops in FF-PUFs. Some methodologies leverage `Explicit' knowledge to construct precise models aligning with the PUFs. A majority employ `Implicit' information for model creation, asserting that any alterations to their model would render the attack ineffective. Consequently, these models often falter in real-world applications. One might argue that iterating through all potential models is an alternative, but even with a comprehensive catalog of PUF designs and specific attacks, the uncertainty of which PUF to initially target makes this approach inefficient. It either results in trial and error or necessitates a deep learning approach demanding considerably more training data. As previously stated, in realistic settings, discerning the exact type of PUF a device employs, even with a uniform authentication protocol, presents a challenge for adversaries. For example, the same 64-stage 6 XOR-APUF in \cite{Reference4} was reported to need 20 minutes to train an attack model. Of course, the difficulty depends on available computing resources, but repeatedly training and testing models until the right one is found is costly. The alternative is an entirely black-box deep learning attack, which infers both the PUF model and the values of a given PUF. The core motivation of this work is to explore the lower bound of viable attackers. To this end, we examine techniques for training PUF prediction models in parallel and without prior design knowledge (as shown in Fig. \ref{fig:motivation}). This is an example of a single expert tool that a low-capability attacker could use to probe for modelable PUFs. We also look at the lower bound of CRPs needed to achieve a high chance of collision between actual and predicted CRPs. The aim is to provide insight into the real risk level for PUF-based authenticated devices against opportunistic, low-effort attackers.  Beyond the scope of practical adversary situations, we've observed a research schism. The community is divided between crafting modelling-resistant PUFs and designing machine learning attacks against them, sparking debates over their security robustness. This divide often inflates concerns and fosters undue confidence in PUF security, primarily due to the absence of a standardized tool to evaluate a PUF's security. In discussions surrounding newly introduced PUFs, disputes arise over their relative security. While unique attack strategies may exist for each, they might employ varying techniques, such as neural networks or evolution strategies. Finding an equitable comparison metric proves challenging. We aim to introduce a universal framework capable of modelling any delay-based PUFs without necessitating setting alterations. In this manner, the requisite number of CRPs can serve as a definitive standard for assessing security levels, thus the quantitative index of modelling resistance of a PUF design.  On the other hand, managing multiple models for multiple PUF-embedded devices could be challenging, where an adversary needs to train, store and load individual PUF models for a target device. 
Thus, to address this issue, in this research, we aim to build a generic PUF model where an attacker needs to maintain only a single model for all PUFs. To summarize, our motivation for this article can be specified through the following research questions (RQs):
\begin{enumerate}
    \item \textbf{RQ 1:} \textit{Can we perform a modelling attack on a PUF without knowing its architectural topologies information? }
    \item \textbf{RQ 2:} \textit{Is it possible for an attacker to maintain \textbf{a single model} to attack multiple cross-architectural PUFs? And would no update of the constructed single model even be required to perform the attack on other PUFs?}
    \item \textbf{RQ 3:} \textit{Considering the practical concerns of training efficiency, time cost and model loading problem, how realistic is it to attack multiple PUFs together with one single model?}
\end{enumerate}

\subsection{Contributions}
Existing literature posits two primary ML attack methods against PUFs: mathematical-model-based and deep-learning-based attacks. Attacking using a predetermined mathematical model is highly efficient. However, it requires an attacker to know the type of PUF and details of its specific implementation (e.g., the number of stages in a delay PUF, the number of XORs in an XOR PUF). On the other hand, PUF inference using deep learning can attack any PUF without knowing as much detail as required by mathematical-model-based methods but requires large sets of training data and some high-level implementation details. In this work, we propose a generic model which does not require any information about the targeted PUF except the CRPs. Further, we expand the model for multi-task learning that could be applied to PUF attacks to exploit commonalities in structure and logic across PUF instances, thereby creating an attack which is generic (like deep learning inference), with reasonable training data and computation costs (like using a predefined model), minimal expertise required to use, and which retains high prediction accuracy. The specific contributions are:

\begin{enumerate}
    \item We define a pragmatic PUF attack scenario based around determining the \textbf{``Minimum Viable Adversary"}. That is the bare minimum of skill, knowledge, and resources required to compromise the authentication system under test.
    \item We propose a \textbf{generic} framework based on multi-experts collaborative learning for modelling delay-based PUFs with minimal knowledge and resources. This represents the kind of tool that significantly lowers the bar for what constitutes a Minimum Viable Adversary. 
    \item We provide a replication of several proposed ML attack methods found in the existing literature and compare their performance to our proposed tool using both simulated and in-silicon PUF data in order to evidence our claim that attacks using predefined mathematical models are non-transferable. 
    \item We demonstrate a successful attack using our tool on the heterogeneous XOR Feed-Forward Arbiter PUF. To the best of our knowledge, this is the \emph{first} published attack against this design.
    
    \item We conduct cross-architectural experiments and show that our model can achieve high accuracy among different kinds of delay-based PUFs without any modifications to the model. For instance, we attack five distinct delay-based PUFs, four unique 4-XOR-APUFs, five different FF-XOR-APUFs and Interpose PUFs, with all attacks attaining a prediction accuracy over 90\%.
    \item We successfully attack multiple PUFs together in one single model, where the PUFs have different settings and structures. We conduct detailed experiments and achieve accuracy beyond $90\%$ among all the experiments.

    \item We provide all code, FPGA implementations of PUFs and datasets used in this paper for the use of the research community\footnote{\textbf{The code and datasets used in this paper are provided in full for the use of the research community at: }\href{https://github.com/AnonymousAppdx/Generic-Framework-for-Modelling-PUFs}{\textbf{https://github.com/AnonymousAppdx/Generic-Framework-for-Modelling-PUFs}}}. We hope this tool will be useful for those working on variant delay-based PUFs as an ``out of the box" attack to test against without needing a lot of modification or computational resources to run.
\end{enumerate}

\section{Preliminaries}
This section lays the foundation for our discussion by providing an overview of Arbiter PUF and its various components. We delve into the intricate world of modelling attacks aimed at PUFs, highlighting representative methods that challenge their security. Furthermore, we introduce the cutting-edge Multi-gate Mixture-of-Experts Model (MMoE) as a promising solution. Lastly, we define the adversary model as the basis for our analysis.

\subsection{Delay-based PUFs}
\label{sub:PUF}
\begin{figure}[t!]
\begin{center}
\makebox[0pt]{\includegraphics[width=90mm]{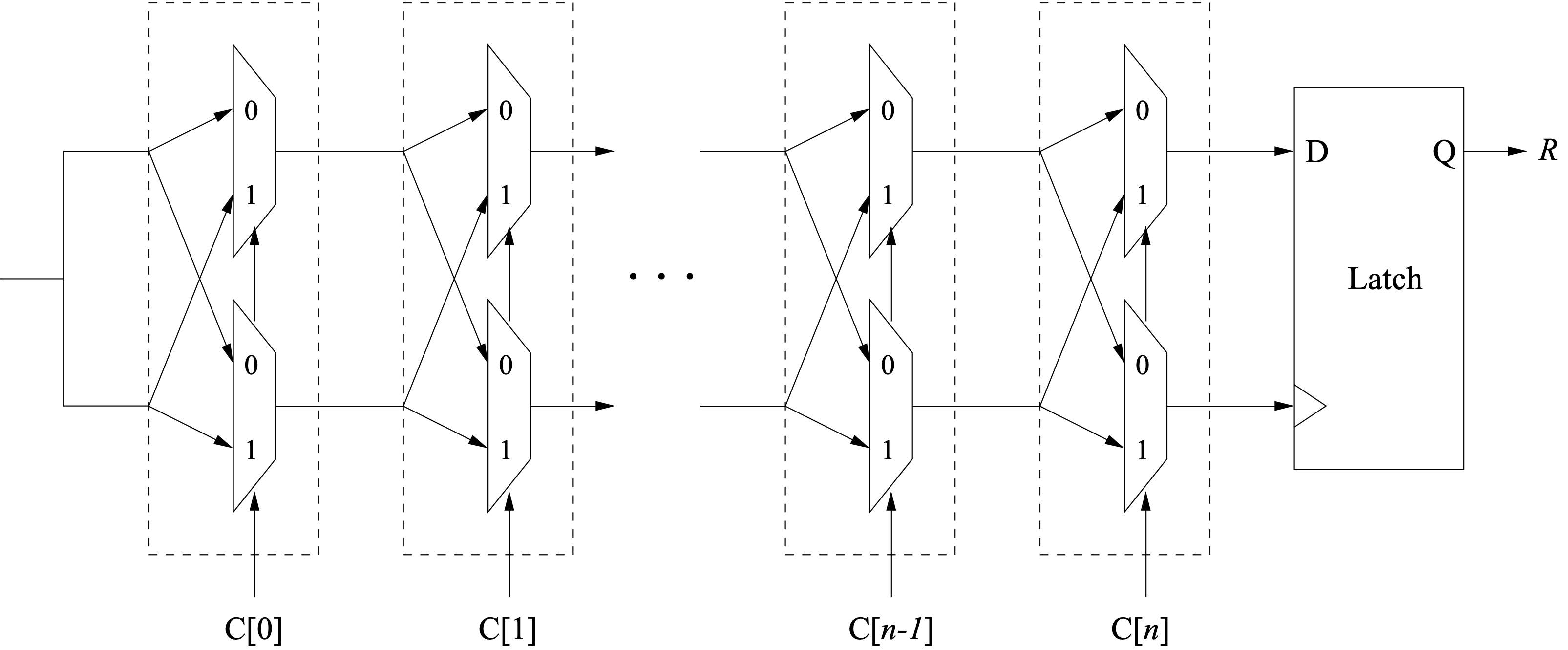}}
\end{center}
\caption{Arbiter PUF delay chain diagram.}
\label{fig:apuf}
\end{figure}

As described in Section \ref{sec:intro}, PUFs can be classified into two distinct categories: Strong PUFs and Weak PUFs. This classification does not reflect their security levels but rather hinges on the number of Challenge-Response Pairs (CRPs) they support.
The exponential number of CRPs in Strong PUFs renders them the preferred choice for device authentication. Their inherent property of random deviation in hardware manufacturing grants them an unclonable physical structure, making it challenging for attackers to model their behaviour. However, they remain susceptible to modelling attacks facilitated by Machine Learning (ML) techniques. In these attacks, an adversary collects a significant number of CRPs and constructs a mathematical model, allowing them to predict the patterns of CRPs accurately. This section aims to elucidate the concepts behind various types of Strong PUFs. An example of an implementation of a strong arbiter PUF is shown in Figure \ref{fig:apuf}. This design exploits the manufacturing variability in the gate delays for randomness and establishes a race condition in a symmetric circuit. The circuit splits an input edge to two multiplexers and creates two identical paths to the output latch based on the input challenge bits $X[0], \cdots, X[n]$. Though the two paths and their propagation times are identical as designed, the random manufacturing variability in the gate delays at the multiplexers will result in one edge arriving first at the latch, with the latch acting as the "arbiter". The figure shows one output (response) bit that depends on the $n$ challenge bits, and multiple such circuits may be used in parallel to obtain additional response bits.\par

\subsection{Classical Machine-Learning Modeling Attacks on PUF}

ML-MA on PUFs have been a significant pain point when developing Strong PUFs, almost since their conception. ML-MA are carried out by adversaries collecting a subset of CRPs from an individual PUF's total CRP space to use as an input to a sophisticated ML algorithm, such that a mathematical model can be generated which learns correlative properties between different challenges and responses. Over the years, almost all Strong PUFs proposed are vulnerable to ML-MA using many different types of ML algorithms, ranging from traditional ML, which is specifically tailored to a given PUF design, to deep learning methods. The first significant work exposing the vulnerability of PUFs to ML-MA was demonstrated by R{\"u}hrmair et al. in \cite{Ruhrmair_modelling}, where the Logistic Regression and Covariance-Matrix Evolutionary Strategy (CMA-ES)  algorithms were exploited to model Arbiter PUFs, Ring Oscillator PUFs, XOR Arbiter PUFs, Lightweight Secure PUFs and Feed-Forward Arbiter PUFs. These attacks required varying numbers of CRPs for training the models, with the simple Arbiter PUFs, at a minimum, requiring simply 640 total CRPs to break the PUF. The more obfuscated PUFs (XOR-Arbiter PUF and Feed-Forward Arbiter PUF), however, generally required many more CRPs before model convergence occurred at up to 500,000 in most cases. While less efficient than traditional ML-MA (on PUFs), deep learning-based modelling attacks can learn latent representation without requiring knowledge of the underlying PUF structure, broadening their use cases \cite{Santikellur_composition, deep_attks_dbl_puf}. More extensively obfuscated APUF designs have shown improved defences against traditional ML attacks; however, feed-forward neural networks (FNNs) have been shown to successfully model up to 5-XOR APUFs, and (4,4)-iPUFs \cite{Santikellur_composition}.


\subsection{Multiple-task Learning}
Multi-task learning is considered to have been an important research topic in the machine learning community for a long time. By transferring the common knowledge shared between different but related tasks, multi-task learning is expected to improve efficiency and model quality on each task. Deep learning has continuously made breakthroughs in various tasks and applications in recent years. Consequently, multi-task learning methods based on deep network architectures have gradually become the research mainstream. The shared bottom network, proposed by Caruana \cite{caruana1997multitask}, is one of the most widely adopted multitask learning methods. It comprises of a shared bottom model structure, where the hidden bottom layers are shared between tasks and several tower networks specific to the task. The shared-bottom model structure enables knowledge transfer among tasks and dramatically reduces the risk of overfitting. Unfortunately, it may suffer from negative transfer as all tasks must utilize the same shared bottom layers, and the differences between tasks are artificially obliterated.

\subsection{ Mixture-of-Experts and Multi-gate Mixture-of-Experts Model}
\label{subsec:MMoE}
Mixture-of-Experts (MoE) layer was first proposed by Robert et al. \cite{moe} as an associative version of competitive learning. By dividing the main tasks into several appropriate subtasks, each of the subtasks can be solved using simple expert networks. The spirit of dividing and learning is suitable for PUF modelling since the basic APUF component can be easily learned using a network with only dozens of neurons. The core of MoE is the gate function, which can be trained to `select' the most suitable experts for the main task. This kind of gate function is usually composed of only several neurons and doesn't cost much resources. It has normalized outputs which are activated by a $Softmax$ function, referring to the weights assigned to each expert. Then each expert will contribute to the final output according to the weights.

\begin{figure}[t!]
\begin{center}
\makebox[0pt]{\includegraphics[width=0.55\textwidth]{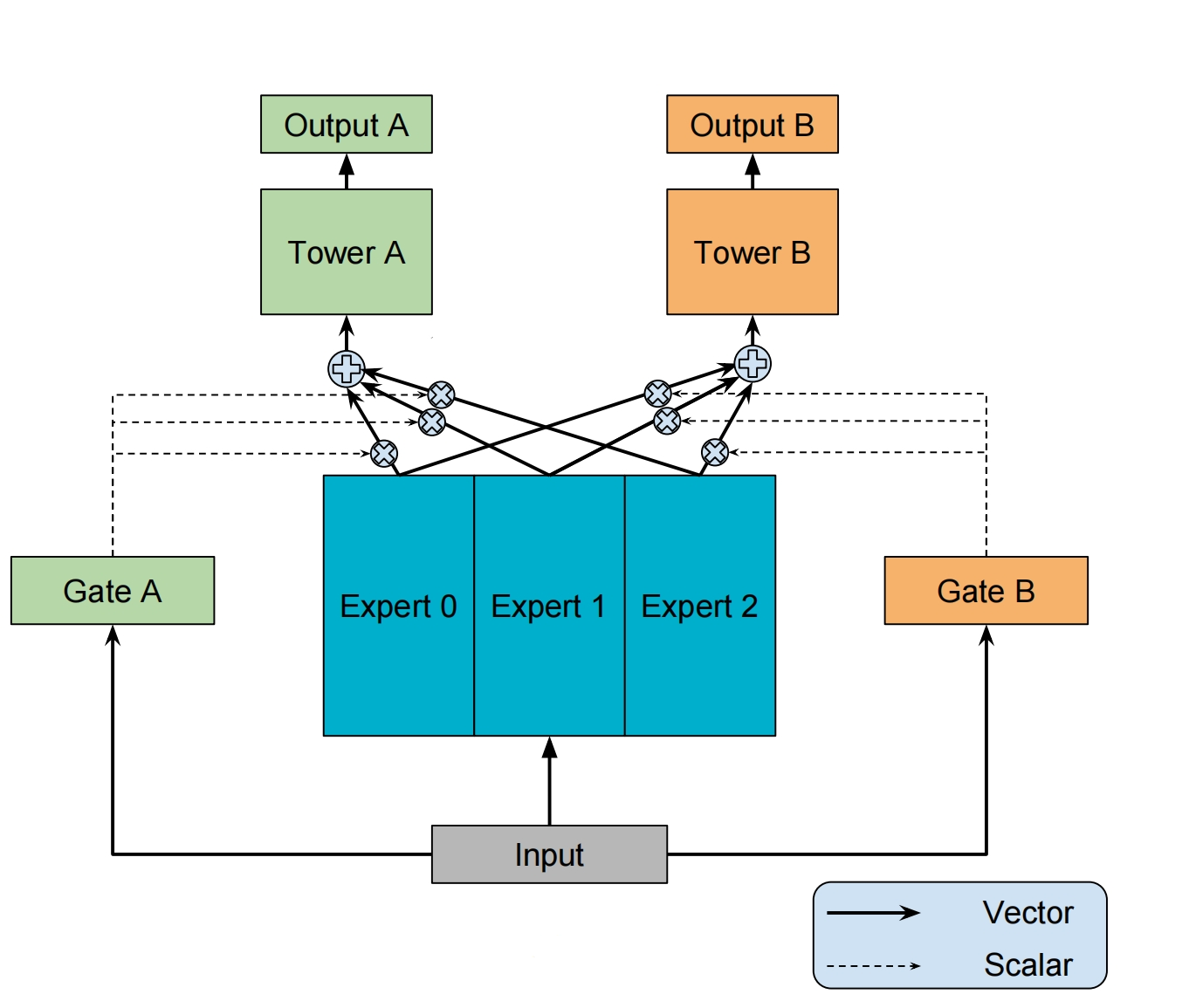}}
\end{center}
\caption{Multi-gate MoE Model}
\label{fig:MMoE}
\end{figure}

The Multi-gate Mixture-of-Experts Model is proposed by Wang et al. \cite{MMoE} to solve the negative transferring produced in multiple tasks modelling. Additionally, it can also capture the similarities across tasks and improve general performance. MMoE adds more gate functions based on the original MoE Model \cite{moe}; the structure is shown in Figure \ref{fig:MMoE}. The core idea of multiple experts is to set up a MoE layer as an ensemble method of multiple individual models. These models are viewed as experts with specific capabilities to solve different tasks. Then, a separate gating network $g^k$ for each task $k$ is used to select suitable experts. The output of task $k$ is:
\begin{equation}
    y_k=h^k\left(f^k(x)\right),
\end{equation}
\begin{equation}
    \text{where } f^k(x)=\sum_{i=1}^n g^k(x)_i f_i(x).
\end{equation}

The gating networks are linear transformations of the input with a softmax layer:
\begin{equation}
    g^k(x)=\operatorname{softmax}\left(W_{g k} x\right).
\end{equation}

\subsection{Minimum Viable Adversary Model}
\label{sec:adversary_model}
In many instances, the threat models of MLAs have not been thoroughly examined or analyzed. This oversight allows adversaries to easily gather excessive information, resulting in an overestimation of the adversary's capabilities and a corresponding underestimation of security robustness. The excessive information includes the type information of PUFs, e.g. XOR-APUFs or FF-APUFs, and the configuration details, such as XOR-APUFs or FF-APUFs, alongside intricate configuration details like the number of XOR gates or the positioning of Feed-Forward Loops. The methodologies for acquiring such information—and the feasibility of doing so—beyond merely Challenge-Response Pairs (CRPs) warrant careful consideration.
Thus the key motivation in this work is to identify the level of actual capability and knowledge an attacker needs in order to pose a threat to a PUF system in real life. In practice, the system designer knows (or at least can know) about PUF vulnerabilities and countermeasures just as much as the adversary. To model a realistic scenario, therefore, we need to incorporate these countermeasures in a way that is still general enough to apply generically to PUF systems. A detailed discussion of various factors we considered can be found in the Appendix. This led us to the following threat model for determining whether a given adversary can perform a viable attack against a realistic PUF network, which uses the following assumptions:

\vspace{5mm}
\textbf{Viable Adversary PUF Threat Model}
\begin{enumerate}
    \item A network of devices exists.
    \item This network comprises of nodes (to be verified) and verifiers.
    \item Nodes are low-resource devices in a low-security environment, and verifiers are high-resource devices in a high(er) security environment.
    \item All nodes contain a PUF. The PUF can be issued challenges and returns a response. 
	\item The PUF has been designed with possible attacks in mind and is as complex as possible within the resource constraints of the system.
	\item There is be a rate limit, \textit{T}, which dictates how often a PUF may be challenged.   
    \item Responses to a given challenge are unique for a given node.
    \item A verifier knows the expected response to any challenge, for any possible configuration, for all nodes it is assigned to verify.
    \item A node is considered authentic (trusted) if it can return the expected response to challenges issued by the corresponding verifier.
    \item The adversary does not have physical access to devices or access to device contents.
	\item The adversary cannot tell which PUF design is being used on any given node but knows which structures are most common in PUF design.
    \item The adversary cannot alter the behaviour of devices. They can trigger any behaviour that is exposed to the network, but doing so risks detection.
	\item There is a time-to-detection after which the network becomes aware of an active adversary and can trigger reconfigurations that reset any modelling progress.
    \item The adversary has a limited window of time to gather CRPs. The number of CRPs which can be gained is a function of the rate limit, \textit{T}, and the worst-case time-to-detection.
    \item If the adversary can predict expected responses under these conditions, they are a \textbf{\textit{Viable Adversary}} and can impersonate a node to gain trusted access.
\end{enumerate}
\vspace{5mm}

Of particular interest in this work is the \textbf{\textit{Minimum Viable Adversary}}, the lowest effort and capability adversary who still has a good chance of success in this threat model. Modelling this kind of adversary gives us insight into the true lower bound of security for a given PUF design. It is of course important to know whether or not a security mechanism is theoretically secure, but even when the answer is "no", it may still provide a useful deterrent. A system which can be broken by 45\% of adversaries can be justifiably called insecure but can equally be viewed as a successful deterrent for 55\% of threats. A security measure's practical value is always relative to how much it costs and how capable the adversaries are. To calculate the actual practical value of adding a PUF to a given system, we need to know how the relative cost of the PUF compares to other lightweight primitives, the properties of the Minimum Viable Adversary, and whether they fall above or below the anticipated capability of real attackers. This is a more complex and situational question than the binary theoretically secure/theoretically insecure dichotomy, but one worth investigating. 

\subsection{What Constitutes a Successful Attack?}
\label{sub:good_attack}
A PUF can be considered `broken' if an adversary can successfully predict complete (bit-perfect) responses with a significant advantage above 50\% accuracy. i.e., random guessing. Intuitively, as per-bit prediction accuracy increases the number of guesses needed for a collision (all bits guessed correctly) reduces. While there is no significant consensus (as of the publication of this work) on a precise per-bit accuracy value at which a PUF is considered vulnerable, a threshold value of 70\% per-bit is commonly applied. The soundness of this is debatable - in the realm of cryptography, even advantages of  50\% +/- 2\% are said to be insufficiently secure for some applications.  In a PUF context, it is something that must be considered in the context of an applied attack. In a scenario where the adversary can perform queries at high speed, the adversary can attempt a brute force prediction analogous to guessing an encryption key. In this case, the assumption that $>$50\% prediction accuracy is a serious flaw is a justified one. If, however, we look at a scenario where the attacker is remote and has to interact with the PUF over a network and via a protocol they cannot modify (the standard assumption for most PUF attack papers) then it becomes trivially easy to deny them that capability. If there is any sort of intrusion detection or time-bounding in play, then the brute force approach becomes much harder to pull off.

For example, a 70\% per-bit prediction accuracy only has a $1.49\times10^{-20}$ chance of a full collision on any given attempt for a 128-bit ID. Something as simple as a 1 nanosecond delay enforced between authentication attempts takes the average time to collide into the order of hundreds of years. Further, while faking authentication once is all well and good, ideally the adversary needs to be able to repeatedly do so without being detected. In order to fake a PUF with a good chance of success in a network that has even the most basic countermeasures, the prediction rate has to be fairly close to the error rate of the actual PUF transmissions. That is, the PUF after on-device error correction has been applied. We suggest that this must be \textit{at minimum} $>$80\% for a 64-bit ID (giving better than 1 in 1 million chance of a full correct guess on any given attempt), and by the same reasoning, 90\% for 128 bits, 95\% for 256 bits, 98\% for 512 bits etc. Within a practical amount of time, less than 80\% accuracy is only viable as an attack if the PUF ID is in the order of 32 bits or less. 



Due to this, in the remainder of this work, we assume a success threshold of 90\% as a reasonable minimum. If a tool could be made which achieves this threshold consistently for any target PUF with an achievable amount of CRPs, then the Minimal Viable Adversary presents a quite serious threat. We believe such a tool is possible, and the following sections will detail an example of one based on the Micture-of-Experts model. Something to emphasise in our general methodology is the goal of consistency - aiming purely for minimal CRPs can result in an attack tool which fails for outlier devices. As such. throughout the following sections, we always use the approach of finding the lowest number of CRPs to achieve the target accuracy, testing for a very large numbers of PUFs, and if a failure is detected, increase the CRPs incrementally until we find no failures. This hopefully captures the limitations of an attack tool intended for use by non-experts.

\section{Proposed Generic Framework}
\label{sec:method}
As discussed in Section \ref{sec:related}, numerous modelling attack methods \cite{Ruhrmair_modelling,Mursi,aseeri2018machine,santikellur2019computationally} have demonstrated commendable prediction accuracy in PUF responses. A shared characteristic among these methods is their reliance on the type or structure of the target PUF as foundational information for modelling. Typically, traditional machine-learning-based models are intricately tailored for a specific PUF type, exhibiting prowess in predicting unacquired CRPs with remarkable precision. Beyond the issue of generality, as delineated in the adversary model in Section \ref{sec:adversary_model}, acquiring such information covertly within a network is impractical. Expecting an adversary to test every possible structure while targeting a PUF is also illogical, given the varying amounts of training data required for different PUFs to orchestrate a successful attack. Considering the `Minimal Viable Adversary', we introduce a generic framework to address these challenges of generality and absence of PUF information in PUF modelling. Here, a singular neural network structure is employed to model diverse PUF types, irrespective of prior PUF type knowledge. An extended version is also proposed to perform multiple-PUF modelling on the combinations of several of the same or different PUFs. As discussed in Section. \ref{sec:related}, many types of delay-based-Strong PUFs share structural similarities. Therefore, it is intuitive to utilise this structural similarity and capture the relationship across these differing PUF instances during modelling when preparing an ML-MA.

\subsection{Generic Framework for a Single PUF}
\label{sub:generic_single}

\begin{figure}[t!]
\begin{center}
\makebox[0pt]{\includegraphics[width=\linewidth]{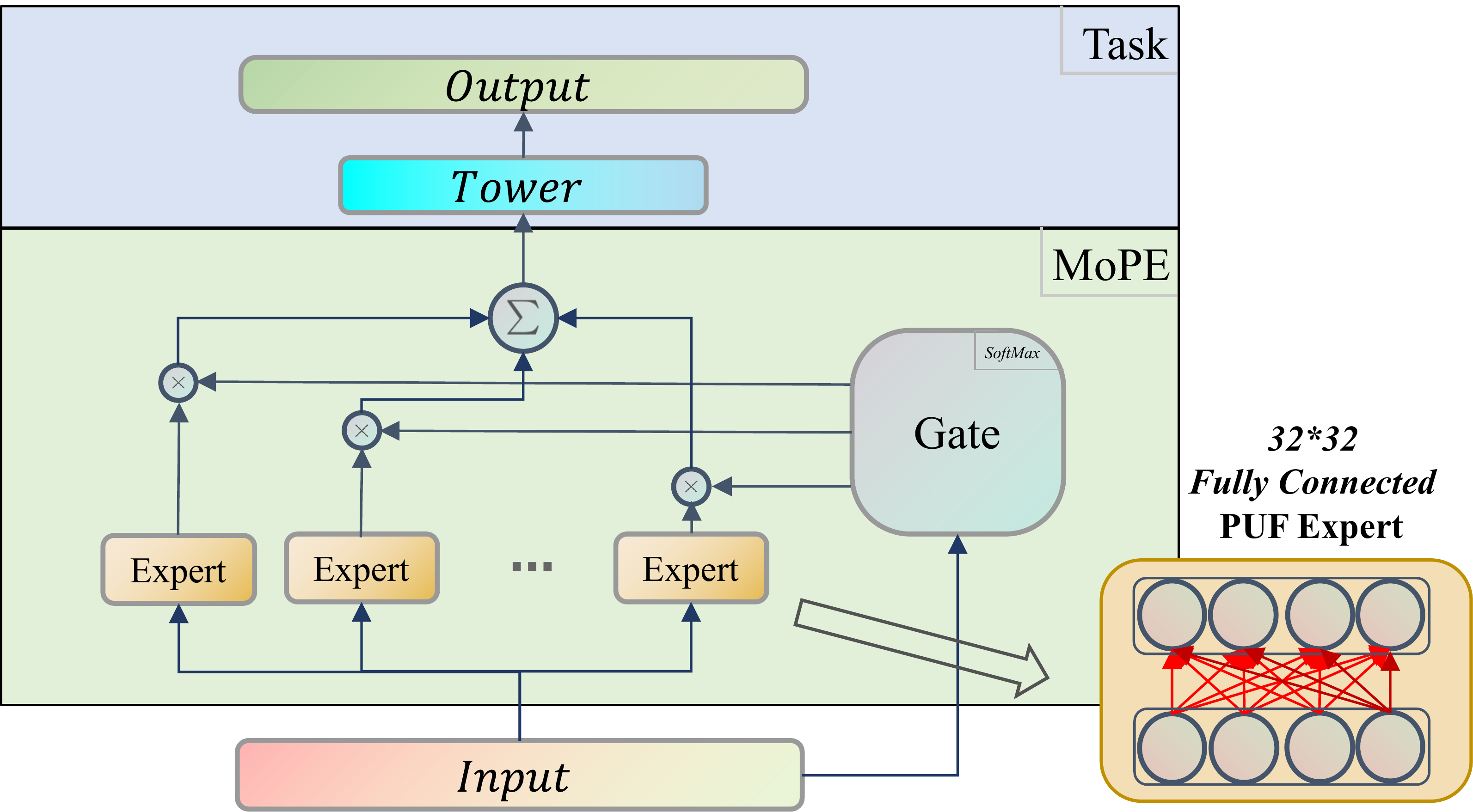}}
\end{center}
\caption{Generic Framework for Modelling a Single PUF.}
\label{fig:moe_1puf}
\end{figure}

We propose Mixture-of-PUF-Experts layer (MoPE) on the basic MoE structure \cite{MMoE}, as briefly outlined in Section \ref{subsec:MMoE}, for modelling PUFs. The structure of MoPE is depicted in Figure \ref{fig:moe_1puf}. The model accepts a challenge as input and produces the predicted response as output. Challenges are processed by an input layer connected to three experts. These experts are tailored to handle the distinct features of CRPs. Each expert comprises of two hidden layers, each with $32$ neurons, we call them `PUF Expert' which is designed for PUF modelling tasks. The first layer is directly connected to the input layer, while the second links to the gate function. The gate function assigns weights to the experts, amalgamates their outputs, and channels them to the tower. 
Initially, we convert the challenge bits $C^{N}$ (where $N$ represents the PUF stages) into the feature vector $X^{N}$, aligning with the structure of delay-based PUF:
\begin{equation}
x_{i}=\prod_{j=i}^{n}c_{j}.
\end{equation}
This transformation aids the model in perceiving the decision boundary as a hyperplane. The response $r$ serves as the label and is adjusted to the range ${0,1}$, if not already within it, to align with the activation function.
Post-feature engineering, the input layer is structured to accommodate these features. In the MoPE layer, we establish $K$ experts, with the count being adaptable based on the number of tasks. The expert structure remains consistent across all PUF types, as two hidden layers equipped with non-linear activation functions are believed to model any function, given sufficient parameters. The $k-$th expert, denoted as $f_{k}(\cdot)$, is designed to extract specific insights or features from the input. Each expert delivers their unique interpretation of the input: $h_{k}(X)=f_{k}(X^{n})$.

To harness the expertise of various experts without overburdening the model with excessive parameters, we introduce the gate function $g(x)$. This function evaluates the features and determines the weight. We employ the $softmax(\cdot)$ activation function post the $N\times K$ kernel $W_{gk}$ to distribute weights among experts and ensure that the model prioritises the most apt one. Consequently, weights are computed as: $g(X)=softmax(W_{NK}(X))$. The weight assigned to the $k-$th expert is represented as $g^{k}(X)$, ensuring that $\sum_{k=1}^{K}g^{k}(X)=1$. Here, we use multiple (5 for single-task, adaptive for multiple-tasks) same experts to ensure the success of modelling. However in most cases, the task won't use all the experts. 

Now, to accelerate the training process, we propose a method called `Sparse Softmax', as shown in \textbf{Algorithm \ref{alg:Sparse_Softmax}}. Sparse Softmax function automatically sets the weights below certain threshold $\tau$ to zeros, which has two benefits. First, the model pays more attention to the more suitable experts and helps accelerate the training. Second, the model can be flexible regarding scale size such that even overdoses of experts won't slow the training down too much, since the backward propagation will cost much more resources and training time than forward propagation, and zero weights need no calculations. The Sparse Softmax function is crucial for the generic framework, since it prevents the overfitting problem for PUFs with simple structures and help the convergence speed when modelling complex PUFs. Subsequently, the MoPE layer's output is derived by amalgamating the outputs of the experts: $mope(X)=\sum_{i=1}^{K}g^{i}(X)h_{i}(X)$. We then establish the tower layer, $T(\cdot)$, tasked with processing the composite information supplied by the experts. This layer then connects to the output layer, which employs the $sigmoid(\cdot)$ activation function to restrict the prediction output to the range ${0,1}$. Our rationale for selecting a dual-layer hidden structure aligns with the perspective of Wisiol et al. in \cite{wisiol2022neural}. Viewing neural networks as a potent instrument for PUF modelling, we are confident that, given ample parameters and layers, PUFs can be effectively modelled, barring optimization constraints. Fewer layers expedite model convergence. Additionally, the MoPE structure's inherent flexibility allows the gate function to integrate multiple experts, facilitating network scalability to accommodate the diverse complexities inherent in PUFs.

\begin{algorithm}
\caption{Sparse Softmax Activation}
\label{alg:Sparse_Softmax}
\begin{algorithmic}[1]
\Require Input vector $\mathbf{W_{NUM_{Experts}}}$, threshold $\tau=0.0001$
\Ensure Sparse softmax vector $\mathbf{\widehat{W}_{NUM_{Experts}}}$
\Procedure{SparseSoftmax}{$\mathbf{W}$, $\tau$}
    \State $\text{sum} \gets 0$
    \For{$j = 1$ \textbf{to} $NUM_{Experts}$}
        \State $\text{sum} \gets \text{sum} + e^{\mathbf{W}_j}$
    \EndFor
    \For{$i = 1$ \textbf{to} $NUM_{Experts}$}
        \State $\mathbf{W}_i \gets \frac{e^{\mathbf{W}_i}}{\text{sum}}$
        \If{$\mathbf{W}_i < \tau$}
            \State $\mathbf{W}_i \gets 0$
        \EndIf
    \EndFor
    \State \Return $\mathbf{\widehat{W}}$
\EndProcedure
\end{algorithmic}
\end{algorithm}



\subsection{Generic Framework for Multiple PUFs}
\label{subsec:method_mul}
\begin{figure}[t!]
\begin{center}
\makebox[0pt]{\includegraphics[width=1.04\linewidth]{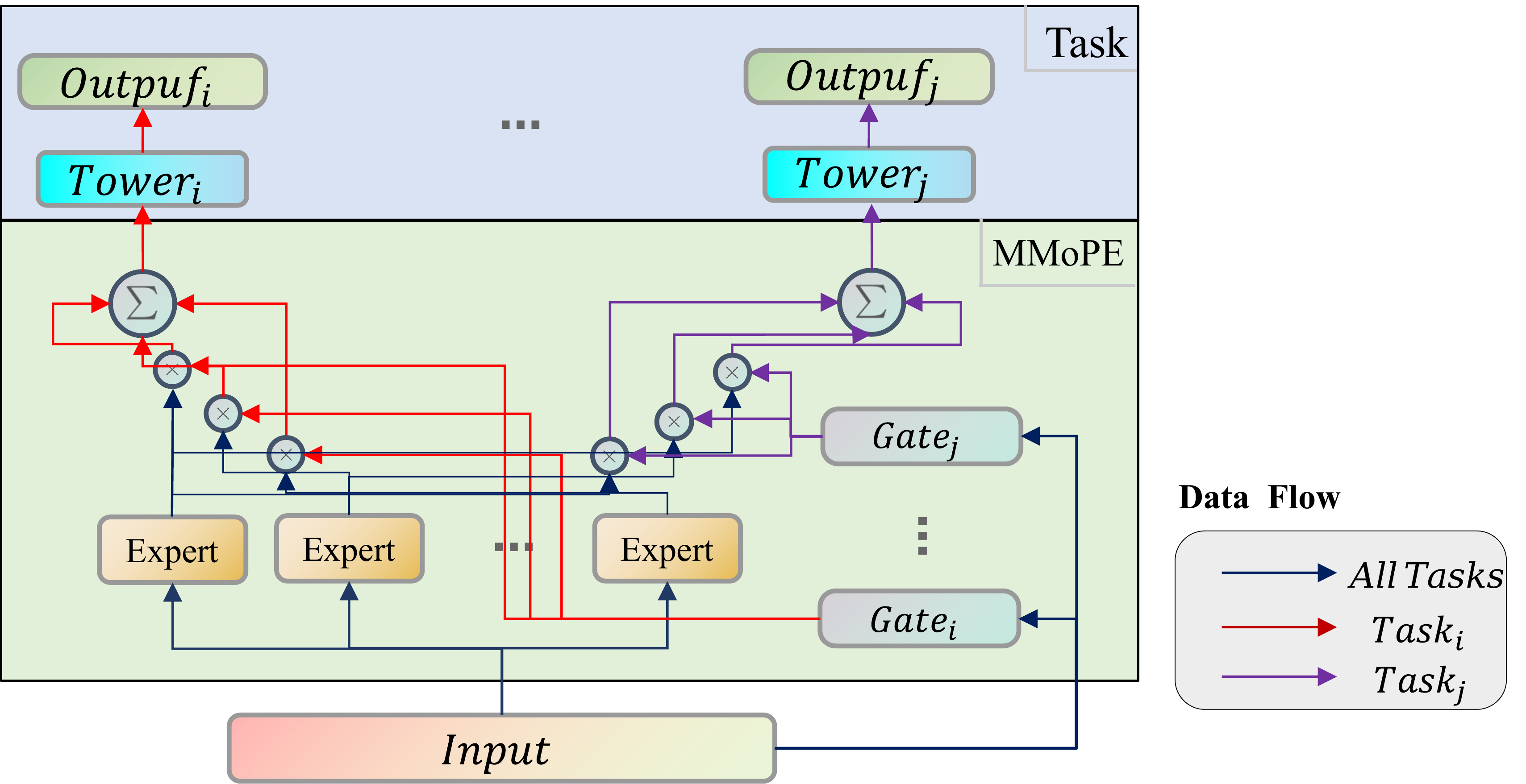}}
\end{center}
\caption{Generic Framework for Modelling Multiple PUFs.}
\label{fig:moe_mpuf}
\end{figure}

This section extends the framework to a multiple-task learning model to enable the attack over multiple unique PUF instances. As discussed in Section \ref{sub:PUF}, the additive delay function represents the inner interaction of PUFs. We can find that they share similar mathematical formulations for the same category of PUF. Taking XOR-Arbiter-PUF as an example, for a $k$-XOR APUF, the responses can be represented as $f(c)=\operatorname{sgn}( \prod_{l=1}^k\left(\left\langle W_l, x\right\rangle+b_l\right))$, signifying that various modelling tasks exhibit commonalities as they employ an identical mathematical representation with varying parameter (delay) values. Thus, a multiple-tasks framework can help improve the modelling performance. The structure of the generic framework for multiple PUFs is shown in Fig. \ref{fig:moe_mpuf}. 

The number of PUF experts can be customized according to the potential types and numbers of PUFs. We add three more PUF experts per extra tasks. The idea is that different structures or numbers of neurons in hidden layers are suitable for different types of PUFs. For some simple PUFs, if the model is too complex, it will be hard for the model to converge without a large amount of training data. This can also be concluded from the comparison results shown by Wisiol et al. in \cite{wisiol2022neural}. In this case, the gate function helps choose the suitable expert from the MoPE layer to alleviate optimization problems caused by too large a model. In Section \ref{sub:generic_single}, we have already shown how to build the single-task model based on MoPE. For multiple tasks, we assign one gate function, tower and output layer to each task. In Fig. \ref{fig:moe_mpuf}, we give an example of modelling several PUFs simultaneously. The feature engineering and input layer remain the same as described in Section \ref{sub:generic_single}. The PUF expert structure is also the same, but the number of PUF experts may differ according to the number of PUFs.

There are $N_{PUFs}$ gate functions, tower functions and output layers for $N_{PUFs}$ PUF models, so-called as $N_{PUFs}$ tasks. For the $t$-th task, the weighed output of MoPE layer is:
\begin{equation}
    moe^{t}(X)=\sum_{i=1}^{K}g^{i}_{t}(X)h_{i}(X).
\end{equation}
Then, the specific tower $T_{t}(\cdot)$ for task $t$ will deal with the extracted features provided by the experts and forward the results to the output layer to give the prediction. In Figure \ref{fig:moe_mpuf}, the red arrows represent dataflow for $Task_{i}$ and the purple ones for $Task_{j}$. The dark blue arrows are the common dataflow for both tasks.

\subsection{Modelling Setup}
We implement our method and replicate other methods using Python 3.8 and TensorFlow 2.4 \cite{abadi2016tensorflow} back-end executed on a Windows laptop with 48 GB of main memory, 5GHz i9-12900H Intel(R) Core processor and NVIDIA GeForce RTX 3070 Ti Laptop GPU. The proposed method should to be generic for any composition of Arbiter-based PUFs; thus, we set up one model and experimented with this model with the same settings. In Section \ref{sec:method}, we introduced the structure of the generic framework; we show the hyperparameters used in the experiments in Table \ref{tab:hp}. We use $Relu$ as the activation function for all the hidden layers, $Softmax$ for the gate function and $Sigmoid$ for the tower function. For all kinds of PUFs, the MoPE has four experts, each with two fully connected hidden layers with $32$ neurons. We adjust the batch size according to the scale of the training dataset. With the number of CRPs denoted as $N_{crp}$, we set $batch\_size=min\{N_{crp},20000\}$. From the experiments, we find that dynamic adjustment can help the network to converge faster. All the codes, data and implementation details are presented in our anonymous GitHub repository\footnote{\textbf{The code and datasets used in this paper are provided in full for the use of the research community at: }\href{https://github.com/AnonymousAppdx/Generic-Framework-for-Modelling-PUFs}{\textbf{https://github.com/AnonymousAppdx/Generic-Framework-for-Modelling-PUFs}}}\label{fnlink}.
\begin{table}
    \centering
    \begin{threeparttable}
        \caption{Hyperparameter Value Used}
    \label{tab:hp}
    \begin{tabular}{cc}
        \toprule
        \textbf{Hyper Parameters} & \textbf{Values} \\
        \midrule
        Kernel Initializer & Glurot uniform \\
        Optimizer & Adam\cite{kingma2014adam} \\
        Hid. Lay. activ. & Relu \\
        Learning rate & Adaptive \\
        Loss function & BCE \\
        \bottomrule
    \end{tabular}
        \end{threeparttable}
\end{table}

\subsection{Data Preparation}
 We conduct experiments on three unique datasets: one simulated dataset built by an additive-delay model, one collected from PUF designs synthesised in hardware, provided by Mursi et al. \cite{Mursi}, and one biased dataset collected from a non-layout-optimized design by ourselves. Each dataset simply consists of CRPs. In our simulated dataset, the challenges are generated using the \textit{pypuf} \cite{pypuf} python library's random generator function as a set of binary strings of length $n$ corresponding to the PUF challenge length. Responses are generated by applying each challenge to each tested PUF instance, each of which is generated with a random seed (for each unique PUF). The seed is randomly chosen and guaranteed to differ for every simulated data generated in this article. For more details, please refer to the code\footnotemark[2].


\section{Generic Framework for Modelling Single PUF}
\label{sec:single}
In this section, we first show how to use the generic framework for any kind of delay-based PUF without knowing any extra information other than CRPs \textbf{and answer \textbf{RQ 1 and RQ 2}}. Then, we present the results and analysis based on the experiments. For our initial comparison, we present the experiment settings and results for the proposed generic framework of a single PUF.

\subsection{Modelling Accuracy Results on Simulated Dataset}

 As shown in Table \ref{tab:results_single}, we present the performance of our method on modelling XOR PUF and compare them with other state-of-the-art machine learning models. Overall, our method successfully performs attacks on different structures of Arbiter-based PUFs without the need to change settings. With the benefit of a fixed structure, rather than a structure that expands as the PUF structure becomes more complex, the training time does not increase exponentially with the complexity of PUFs. From the accuracy results, we can find that our method does not require more training CRPs to counteract the negative effects brought by the fixed network structure.
During the experiments, the strategy for every method is to begin the modelling at a reasonable amount of CRPs; then, if the attack succeeds with accuracy beyond $90\%$, we attempt more random PUFs. If an attempt fails, we increase the amount of CRP training data; otherwise, if the method models all the PUFs, we decrease the amount of training data and continue the test. The idea is to find the point where each attack succeeds enough to be \textit{practically useful}, as described in Section \ref{sec:adversary_model}, against even fairly large PUFs, can achieve this consistently, and uses as few CRPs as possible within those constraints.

\begin{table}[ht]
    \centering
    \begin{threeparttable}
            \caption{Modelling results of for Single XOR PUF on Simulated CRPs}
        \label{tab:results_single}
        \begin{tabular}{cccccc}
            \toprule
            \textbf{Method} & \textbf{k} & \textbf{crp} & \textbf{time} & \textbf{acc} & \textbf{Memory} \\
            \midrule
            \textbf{Rührmair et al}\cite{Ruhrmair_modelling}& \multirow{3}{*}{2} & 8k & \textless20sec & \textgreater 97.0\% &  1.78 GiB \\ 
            \textbf{Aseeri et al}\cite{aseeri2018machine}&  & 8k & \textless20sec & 97.0\% &  1.79 GiB \\ 
            \textbf{Mursi et al}\cite{Mursi}&  & 8k & \textless1min & \textgreater98.0\% &  1.86 GiB \\
            \textbf{Proposed Scheme}& & 8k & \textless 20sec & $>$94.0\% & 2.06 GiB \\
            \midrule
            \textbf{Rührmair et al}\cite{Ruhrmair_modelling}& \multirow{3}{*}{3} & 20k & \textless20sec & \textgreater99.0\% &  1.84 GiB \\ 
            \textbf{Aseeri et al}\cite{aseeri2018machine}&  & 24k & \textless20sec & \textgreater97.0\% & 1.81 GiB \\ 
            \textbf{Mursi et al}\cite{Mursi}&  & 24k & \textless1min & \textgreater98.0\% &  1.88 GiB \\
            \textbf{Proposed Scheme}&  & 24k & \textless1min & $>$95\% & 2.15 GiB \\
            \midrule
            \textbf{Rührmair et al}\cite{Ruhrmair_modelling}& \multirow{3}{*}{4} & 30k & \textless20sec & \textgreater99.0\% &  1.91 GiB \\ 
            \textbf{Aseeri et al}\cite{aseeri2018machine}&  & 100k & \textless1min & \textgreater98.0\% & 1.91 GiB \\ 
            \textbf{Mursi et al}\cite{Mursi}&  & 120k & \textless1min & \textgreater98.0\% &  1.97 GiB \\
            \textbf{Proposed Scheme}& & 80k & \textless1min & $>$97.0\% & 2.26 GiB \\
            \midrule
            \textbf{Rührmair et al}\cite{Ruhrmair_modelling}& \multirow{3}{*}{5} & 260k & \textless20sec & \textgreater99.0\% &  1.88 GiB \\ 
            \textbf{Aseeri et al}\cite{aseeri2018machine}&  & 400k & \textless2min & \textgreater95.0\% &  2.32 GiB\\ 
            \textbf{Mursi et al}\cite{Mursi} &  & 240k & \textless1min & \textgreater98.0\% &  2.29 GiB \\
            \textbf{Proposed Scheme}& & 240k & \textless1min & \textgreater97.0\% & 2.24 GiB \\
            \midrule
            \textbf{Rührmair et al}\cite{Ruhrmair_modelling}& \multirow{3}{*}{6} & 3M & \textless1min & \textgreater99.0\% &  2.34 GiB \\ 
            \textbf{Aseeri et al}\cite{aseeri2018machine}&  & 1.6M & \textless2min & \textgreater99.0\% & 4.29 GiB \\ 
            \textbf{Mursi et al}\cite{Mursi}&  & 1.6M & \textless2min & \textgreater99.0\% &  3.04 GiB \\
            \textbf{Proposed Scheme}& & 800k & \textless2min & \textgreater96.0\% & 3.31 GiB \\
            \midrule
            \textbf{Rührmair et al}\cite{Ruhrmair_modelling} & \multirow{3}{*}{7} & 20M & \textless1hr & \textgreater98\% &  4.61 GiB \\ 
            \textbf{Aseeri et al}\cite{aseeri2018machine}&  & 5M &  \textless20min& \textgreater97\% & 8.05 GiB \\ 
            \textbf{Mursi et al}\cite{Mursi} & & 4M & \textless20min & \textgreater98.0\% & 8.07 GiB\\
            \textbf{Proposed Scheme}& & 2.4M & \textless20min & \textgreater98.0\% & 5.08 GiB \\
            \bottomrule
        \end{tabular}
    \end{threeparttable}
\end{table}

\subsubsection{XOR Arbiter PUF}
We mainly refer to two state-of-the-art neural network models \cite{aseeri2018machine,Mursi} and one logistic regression model \cite{Ruhrmair_modelling} to compare our results. They offer detailed guidance for implementing the codes that can be easily replicated. Additionally, their claimed accuracy is close to our observations. To make the comparison fair and reasonable, we evaluate the performance of compared models by reproducing them using the same dataset and hardware resource. Besides our strategy for finding a stable amount of training data, we tried to find the best hyperparameters for compared schemes that were not claimed, e.g. the training batch size.

For the method presented by Rührmair et al. \cite{Ruhrmair_modelling}, we use code provided by Wisiol et al. in the PyPuf Python library \cite{pypuf}. This attack builds the network strictly according to the mathematical models. For small stages of XOR-APUFs, it outperforms all other methods in training time (lower is better) and accuracy. As stated in \cite{Nguyen_iPUF} \textbf{Theorem 1}, the logistic regression method is the most powerful attack among classical machine learning attacks; however, as the number of stages increases, LR consumes the most training data and time. \textit{Intuitively, when the number of XORs is more than 5, our proposed method has more advantages.}

For the attacks presented by Asseri et al. in \cite{aseeri2018machine} and Mursi et al. in \cite{Mursi}, we implement their methods using the same setting claimed in their paper, including the kernel initializer, optimizer, learning rate, activation functions, and loss function. However, we optimize the batch size and adapt it according to the scale of training data. During the difficult reproducing work of different schemes, we find that the success of an attack highly relies on many factors: the dataset, the initial state of the model, the structure of the network, and even the batch size. Many ML-MA works indicate optimal accuracy and present their best accuracy on specific PUFs with few training CRPs. When we try to apply the method to a different dataset, it fails or can not achieve a reliable success rate. In some cases, different initialisations of the kernels succeed on different datasets and fail on others. The modelling process can be susceptible to differing hyperparameters. This kind of unreliable attack is unacceptable for realistic adversaries. Besides, methods Asseri et al. \cite{aseeri2018machine}, and Mursi et al. \cite{Mursi} designed different structures for different PUFs. In most cases, they can not learn the PUFs of different types of PUFs, which means they need to know the type information. We discerned their acute sensitivity to the PUF's structure through our attack implementations on these schemes. This underscores the indispensability of type/structure information for mounting successful attacks. For instance, in \cite{Mursi}, the neural network devised for attacking a $k$-XOR Arbiter PUF comprises three fully connected hidden layers sized $\{2^{k-1},2^{k},2^{k}\}$. This implies that for a 5-XOR APUF and 6-XOR APUF, the hidden layer structures should be $\{2^{4},2^{5},2^{4}\}$ and $\{2^{5},2^{6},2^{5}\}$, respectively. We validated their efficacy on the stated CRP quantities, $200$k and $200$M. Yet, when we experimented with the $\{2^{4},2^{5},2^{4}\}$-structure neural network for the 6-XOR APUF and the $\{2^{5},2^{6},2^{5}\}$-structure for the 5-XOR PUF, both attempts were unsuccessful. These outcomes underscore that non-generic models are meticulously crafted; simplistic models falter with complex PUFs, and conversely, intricate models struggle with simpler PUFs. Potential reasons could range from optimization challenges to under-fitting in the former scenario and over-fitting in the latter.

Next, we evaluate the generic capability of different models as shown in Table \ref{tab:comp_gene}, from where we can find that our proposed scheme can achieve good accuracy in modelling different PUFs. Table \ref{tab:comp_gene} shows the performance of different models, including MoPE, Mursi et al. \cite{Mursi}, Aseeri et al. \cite{aseeri2018machine}, Rührmair et al. \cite{Ruhrmair_modelling} and Mishara et al. \cite{mishra2024calypso}, on XOR-PUFs ranging from 2 to 7 XORs. The sub-columns of each method represent the model designed for the specific type of PUF. For example, the cell positioned at \{3-XOR, Mursi \cite{Mursi}, 2\} indicates that the accuracy of modelling 3-XOR-PUF using 2-XOR-PUF-Model is $98\%$. The table shows that our proposed model can perform the modelling attack across any XOR-PUFs with accuracy ranging from $94-98\%$. On the other hand, this kind of cross-architectural modelling capability has not been considered in all other methods. Consequently, this leads to a pertinent question: How can we compare the security levels of two distinct PUFs? Furthermore, how should we select the model and conduct the evaluations? For instance, if we consider the 7-XOR-PUF model proposed by Aseeri et al. and evaluate the 6-XOR-PUF and 5-XOR-PUF, we get the accuracy of $98\%$ and $50\%$, which denotes that security-level of 5-XOR-PUF is stronger than 6-XOR-PUF, which is not true. In a nutshell, if we consider evaluating the security performance of a structure-unknown PUF or comparing distinct PUFs, our proposed model has a significant advantage over all other methods.

\begin{table*}[ht]
\centering
\caption{Comparison of Generic Modelling Capability}
\label{tab:comp_gene}
\resizebox{\textwidth}{!}{
\begin{tabular}{|l|c|*{23}{c|}}
\hline
\rowcolor{gray!50}
Method & MoPE & \multicolumn{6}{c|}{Mursi\cite{Mursi}} & \multicolumn{6}{c|}{Aseeri\cite{aseeri2018machine}} & \multicolumn{6}{c|}{Rührmair\cite{Ruhrmair_modelling}} & \multicolumn{5}{c|}{Mishra\cite{mishra2024calypso}} \\
\hhline{|>{\arrayrulecolor{gray!50}}->{\arrayrulecolor{black}}------------------------|}
PUFs/Model & Ours & 2 & 3 & 4 & 5 & 6 & 7 & 2 & 3 & 4 & 5 & 6 & 7 & 2 & 3 & 4 & 5 & 6 & 7 & 1 & 2 & 3 & 4 & 5\\
\hline
2-XOR & \cellcolorme{94} & \cellcolorme{98} & \cellcolorme{98} & \cellcolorme{98} & \cellcolorme{98} & \cellcolorme{95} & \cellcolorme{95} & \cellcolorme{98} & \cellcolorme{97} & \cellcolorme{96} & \cellcolorme{93} & \cellcolorme{92} & \cellcolorme{93} & \cellcolorme{99} & \cellcolorme{97} & \cellcolorme{99} & \cellcolorme{59} & \cellcolorme{50} & \cellcolorme{52} &
\cellcolorme{75} & - & - &- &-\\
\hline
3-XOR & \cellcolorme{95} & \cellcolorme{98} & \cellcolorme{98} & \cellcolorme{98} & \cellcolorme{98} & \cellcolorme{96} & \cellcolorme{95} & \cellcolorme{98} & \cellcolorme{97} & \cellcolorme{97} & \cellcolorme{95} & \cellcolorme{92} & \cellcolorme{87} & \cellcolorme{66} & \cellcolorme{99} & \cellcolorme{99} & \cellcolorme{99} & \cellcolorme{50} & \cellcolorme{54} &
- & \cellcolorme{78} & - &- &-\\
\hline
4-XOR & \cellcolorme{97} & \cellcolorme{98} & \cellcolorme{98} & \cellcolorme{98} & \cellcolorme{98} & \cellcolorme{98} & \cellcolorme{51} & \cellcolorme{50} & \cellcolorme{50} & \cellcolorme{98} & \cellcolorme{50} & \cellcolorme{56} & \cellcolorme{55} & \cellcolorme{60} & \cellcolorme{63} & \cellcolorme{99} & \cellcolorme{50} & \cellcolorme{99} & \cellcolorme{52} &
- & - & \cellcolorme{77} &- &-\\
\hline
5-XOR & \cellcolorme{97} & \cellcolorme{98} & \cellcolorme{98} & \cellcolorme{98} & \cellcolorme{98} & \cellcolorme{99} & \cellcolorme{50} & \cellcolorme{50} & \cellcolorme{50} & \cellcolorme{50} & \cellcolorme{95} & \cellcolorme{50} & \cellcolorme{50} & \cellcolorme{55} & \cellcolorme{58} & \cellcolorme{52} & \cellcolorme{99} & \cellcolorme{50} & \cellcolorme{50} &
- & - & - &\cellcolorme{80} &-\\
\hline
6-XOR & \cellcolorme{96} & \cellcolorme{50} & \cellcolorme{50} & \cellcolorme{50} & \cellcolorme{50} & \cellcolorme{99} & \cellcolorme{50} & \cellcolorme{50} & \cellcolorme{50} & \cellcolorme{50} & \cellcolorme{50} & \cellcolorme{99} & \cellcolorme{98} & \cellcolorme{50} & \cellcolorme{54} & \cellcolorme{59} & \cellcolorme{67} & \cellcolorme{99} & \cellcolorme{50} &
- & - & - & - &\cellcolorme{79}\\
\hline
7-XOR & \cellcolorme{98} & \cellcolorme{50} & \cellcolorme{50} & \cellcolorme{50} & \cellcolorme{50} & \cellcolorme{50} & \cellcolorme{98} & \cellcolorme{50} & \cellcolorme{50} & \cellcolorme{50} & \cellcolorme{50} & \cellcolorme{50} & \cellcolorme{97} & \cellcolorme{50} & \cellcolorme{50} & \cellcolorme{50} & \cellcolorme{50} & \cellcolorme{50} & \cellcolorme{99} & - &- &- &- &-\\
\hline
\end{tabular}}
\end{table*}

\subsubsection{XOR Feed-Forward Arbiter PUF and Interpose PUF}
In this section, we demonstrate the versatility of our model in adapting to a range of modelling tasks. Specifically, we apply our generic model to various additional delay-based PUFs, encompassing even those PUFs previously resistant to successful attacks. Avvaru et al. introduced the homogeneous and heterogeneous Feed-Forward XOR PUFs in \cite{avvaru2020homogeneous}. Subsequent to their work, a multitude of machine learning models were proposed to target FF-APUFs \cite{wisiol2022neural}. A large portion of these models capitalize on the inconsistent reliability of PUF designs, focusing particularly on homogeneous XOR FF PUFs with uniform loop positions. In contrast, heterogeneous XOR-FF-APUFs are largely considered resilient against modelling attacks. As evidenced in Figure \ref{tab:results_FF_IPUF}, we successfully modelled $2-$loop FF PUFs with 2 XOR stages and $1-$loop FF PUFs with 3 XOR stages, achieving accuracy exceeding $95\%$ and $98\%$, respectively. The Interpose PUF was introduced by Nguyen et al. in \cite{Nguyen_iPUF}. Although it was later targeted using the `divide-and-conquer' technique \cite{Wisiol_split}, our results, as depicted in Figure \ref{tab:results_FF_IPUF}, confirm that our proposed model can adeptly launch successful attacks on various configurations of the Interpose PUF without necessitating any structural alterations to the model itself.
\begin{table}
    \centering
    \begin{threeparttable}
            \caption{Modelling results of for Feed-Forward PUFs and Interpose PUFs.}
        \label{tab:results_FF_IPUF}
        \begin{tabularx}{\linewidth}{p{2.0cm}p{0.95cm}p{0.9cm}p{0.64cm}p{0.7cm}p{0.7cm}}
            \toprule
            \textbf{Type} & \textbf{k} & \textbf{Loops} & \textbf{crp} & \textbf{time} & \textbf{acc} \\
            \midrule
             \multirow{12}{*}{\parbox{1.5cm}{\textbf{Homogeneous}\\ \textbf{FF-PUF}}} & \multirow{5}{*}{1} & 1 & 20k & $<$2min & $>$94\% \\
             &  & 2 & 120k & $<$2min & $>$97\% \\
             &  & 3 & 250k & $<$2min & $>$98\% \\
             &  & 4 & 500k & $<$2min & $>$98\% \\
             &  & 5 & 1M & $<$10min & $>$94\% \\
             \cmidrule(lr){3-6}
             & \multirow{5}{*}{2} & 1 & 90k & $<$2min & $>$93\% \\
             &  & 2 & 200k & $<$10min & $>$97\% \\
             &  & 3 & 400k & $<$10min & $>$93\% \\
             &  & 4 & 800k & $<$20m & $>$85\% \\
             &  & 5 & 1.6M & $<$1hr & $>$90\% \\
             \cmidrule(lr){3-6}
             & \multirow{2}{*}{3} & 1 & 120k & $<$2min & $>$90\% \\
             &  & 2 & 400k & $<$10min & $>$93\% \\
            \midrule
            \multirow{3}{*}{\parbox{1.5cm}{\textbf{Heterogeneous}\\ \textbf{FF-PUF}}} & 2 & 1 & 160k & $<$2min & $>$98\% \\
            & 3 & 1 & 640k & $<$10min & $>$98\% \\
            & 2 & 2 & 400k & $<$2min & $>$95\%\\
            \midrule
            \multirow{4}{*}{\parbox{1.5cm}{\textbf{Interpose}\\ \textbf{PUF}}} & \textbf{Upper chains} & \textbf{Lower chains} & \textbf{crp} & \textbf{time} & \textbf{acc} \\
            & 1 & 5 & 480k & $<$2min & $>$98\% \\
            & 3 & 3 & 320k & $<$3min & $>$98\% \\
            & 4 & 4 & 800k & $<$1hr & $>$96\% \\
            & 5 & 5 & 2M & $<$1hr & $>$96\% \\
            \bottomrule
        \end{tabularx}
    \end{threeparttable}
\end{table}

\subsection{Modelling Accuracy Results on Silicon CRPs}

To avoid faulty evaluations caused by wrongly generated simulated data, we validate our method on both unbiased real-world data provided by Mursi et al. \cite{Mursi} and biased implementations. We include the biased dataset primarily to demonstrate the profound impact even minimal (and potentially otherwise accepted) amounts of bias can have in providing knowledge to the adversary's MLA. We randomly select the demand amount of CRPs from the silicon data and apply our framework without changing any model settings, which is needed for all other models. The results are shown in Table \ref{tab:results_single_real}. We can find the accuracy does not decrease at all for any stage of XOR-APUFs when we use the same amount of CRPs.

\begin{table}
    \centering
    \begin{threeparttable}
            \caption{Modelling results of for Single XOR PUF on Silicon CRPs, from \cite{Mursi} and our non-optimized implementations.}
        \label{tab:results_single_real}
        \begin{tabular}{ccccccc}
            \toprule
            & \multicolumn{3}{c}{\textbf{Mursi et al. \cite{Mursi}}} & \multicolumn{3}{c}{\textbf{Our data}} \\
            \cmidrule(lr){2-4} \cmidrule(lr){5-7}
            \textbf{k} & \textbf{crp} & \textbf{time} & \textbf{acc} & \textbf{crp} & \textbf{time} & \textbf{acc} \\
            \midrule
            4 & 40k & \textless20sec & 92.56\% & 40k & \textless20sec & 92.90\% \\
            5 & 160k & \textless20sec & 95.21\%& 120k & \textless1min & 95.36\% \\
            6 & 560K & \textless1min & 94.56\% & 120k & \textless1min & 95.36\%\\
            7 & 1.6M & \textless1min & 93.38\% & 600k & \textless10min & 95.81\%\\
            \bottomrule
        \end{tabular}
    \end{threeparttable}
\end{table}

\begin{table}[t]
    \centering
    \begin{threeparttable}
            \caption{Modelling results of for Multiple (\textbf{Two}) XOR PUFs on Simulated CRPs.}
        \label{tab:results_multiple_comb}
        \begin{tabular}{ccccc}
            \toprule
            \textbf{Method} & \textbf{k} & \textbf{crp} & \textbf{time} & \textbf{acc} \\
            \midrule
            \textbf{Aseeri et al}\cite{aseeri2018machine}& \multirow{3}{*}{2} & 10k & \textless20sec & 71.5\% \\ %
            \textbf{Mursi et al}\cite{Mursi}&  & 10k & \textless20sec & 67.34\% \\
            \textbf{Proposed Scheme}& & 8k & \textless 20sec & 93.50\% \\
            \midrule
            \textbf{Aseeri et al}\cite{aseeri2018machine}& \multirow{3}{*}{3} & 30k & \textless20sec & 72\% \\ %
            \textbf{Mursi et al}\cite{Mursi}&  & 30k & \textless20sec & 74.39\% \\
            \textbf{Proposed Scheme}&  & 24k & \textless20sec & 93.02\% \\
            \midrule
            \textbf{Aseeri et al}\cite{aseeri2018machine}& \multirow{3}{*}{4} & 100k & \textless20sec & 73.72\% \\ %
            \textbf{Mursi et al}\cite{Mursi}&  & 100k & \textless20sec & 74.60\% \\
            \textbf{Proposed Scheme}& & 80k & \textless20sec & 93.99\% \\
            \midrule
            \textbf{Aseeri et al}\cite{aseeri2018machine}& \multirow{3}{*}{5} & 400k & \textless20sec & 50\% \\ %
            \textbf{Mursi et al}\cite{Mursi} &  & 400k & \textless20sec & 74.89\% \\
            \textbf{Proposed Scheme}& & 240k & \textless20sec & 97.88\% \\
            \midrule
            \textbf{Aseeri et al}\cite{aseeri2018machine}& \multirow{3}{*}{6} & 2M & \textless1min & 74.08\% \\ %
            \textbf{Mursi et al}\cite{Mursi}&  & 2M & \textless1min & 74.08\% \\
            \textbf{Proposed Scheme}& & 800k & \textless20sec & 95.04\% \\
            \midrule
            \textbf{Aseeri et al}\cite{aseeri2018machine}& \multirow{3}{*}{7} & 5M & \textless1hr & 74.20\% \\ %
            \textbf{Mursi et al}\cite{Mursi} & & 5M & \textless20min & 50\% \\
            \textbf{Proposed Scheme}& & 2.4M & \textless20min & 98.04\% \\
            \bottomrule
        \end{tabular}
    \end{threeparttable}
\end{table}

\begin{figure}[ht]
\begin{center}
\makebox[0pt]{\includegraphics[width=\linewidth]{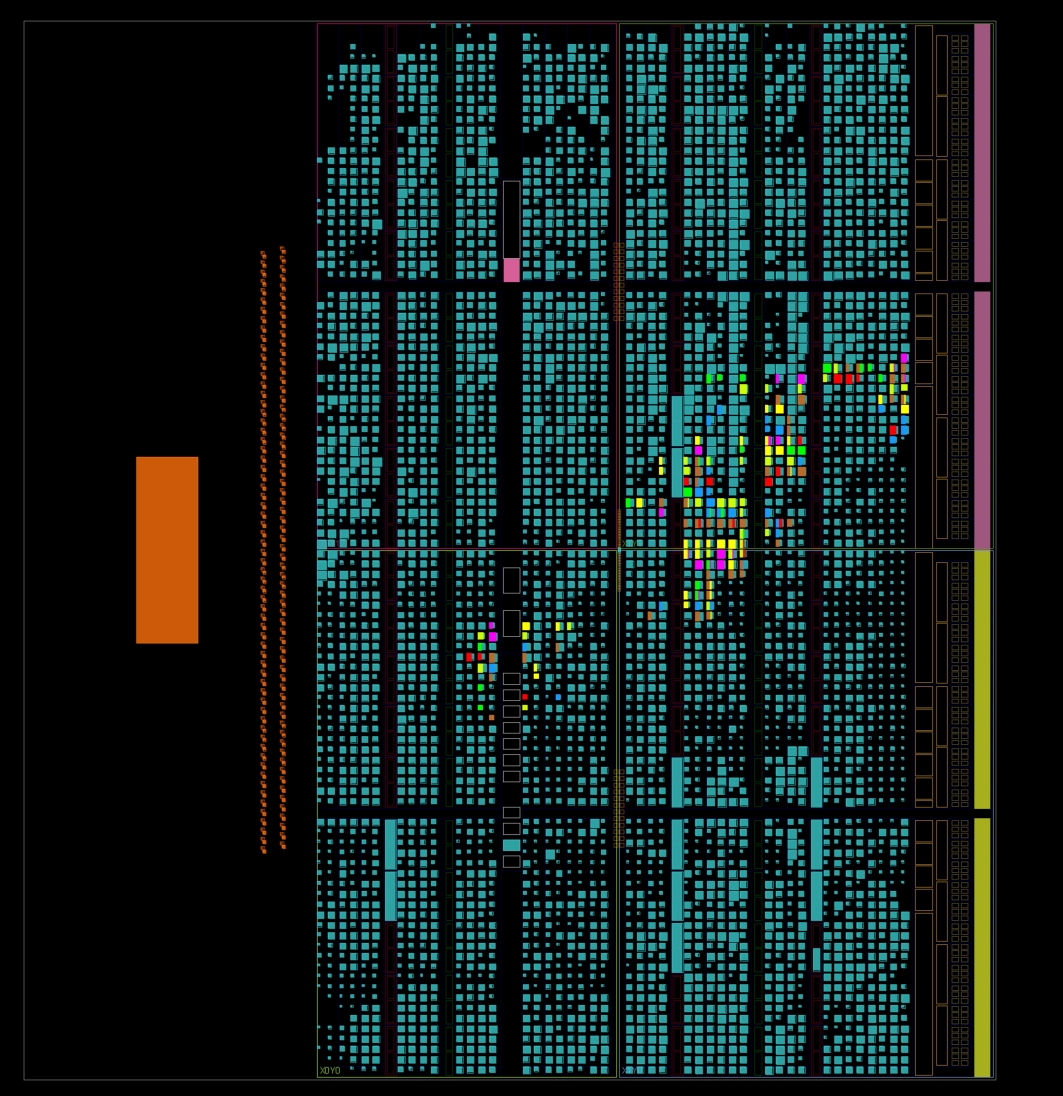}}
\end{center}
\caption{Hardware layout for $7$ PUFs on XC7Z010 FPGA Board. No optimization design for the layout was performed.}
\label{fig:hardware}
\end{figure}

Then, we implement 64-stage Arbiter PUFs on a Zynq-7000 FPGA using the Vivado design suite. Verilog hardware description language is used to build the PUF design. No placement design is applied to these PUFs. The hardware layout is shown in Figure \ref{fig:hardware}. We applied a DRAM controller to transfer challenges and obtain responses between the PL and PS sides. For each single Arbiter PUF, we evaluate its performance. The average bias is around 55\%, which makes the PUF more vulnerable to ML-MA. As shown in Table \ref{tab:results_single_real}, much fewer CRPs are needed to perform successful attacks compared to unbiased implementations. 

\subsection{Resource Consumption and Efficiency Analysis}
In many modelling attack works targeting PUFs \cite{wisiol2022neural, Santikellur_composition, Mursi,aseeri2018machine}, the structure of the model varies for the different PUFs. As our framework is designed for a generic purpose, it is not suitable to compare the scale of models directly for consumption evaluations. However, we list the time cost and memory usage in Fig. \ref{tab:results_single}. The time cost of our proposed scheme is not much different from other compared methods and outperforms Rührmair et al \cite{Ruhrmair_modelling} and Aseeri et al. \cite{aseeri2018machine} when attacking 7-XOR APUF. We used more memory in our scheme for attacking, which is attributed to the number of neurons and layers that are required in our model. For 7-XOR APUF, 5.08 GiB memory is needed, and methods in \cite{Ruhrmair_modelling, aseeri2018machine, Mursi} need 4.61, 8.05, 8.07 GiB. We clarify here that our memory usage might be different than that claimed by the authors; in addition to the hardware differences, we used larger batch sizes to speed up the training. For our model, the memory usage between different tasks mainly depends on the amount of dataset, which determines the batch size. Overall, we fixed the structure for all types of PUFs and the most complex PUF determines the hardware requirement for our model.

\begin{table*}[!ht]
\fontsize{6pt}{7.2pt}\selectfont
\begin{threeparttable}
\setlength\tabcolsep{7pt} 
\centering
\caption{Results of Attacking Different Combinations of Different PUFs with Our Proposed Scheme}
\label{tab:combinations}
\begin{tabular*}{\textwidth}{@{\extracolsep{\fill}} ccccc}
\toprule
\multirow{2}{*}{\textbf{Comb. No.}}& \multirow{2}{*}{\textbf{Combinations}} &\textbf{Numbers } & \multicolumn{2}{c}{\textbf{Performance}} \\
& & \textbf{of PUFs} & \textbf{Average Accuracy} & \textbf{Running time} \\
\midrule
\multirow{4}{*}{\uppercase\expandafter{\romannumeral 1}}& 4 $\times$ 3-\textbf{XOR APUF}\tnote{\textdagger} & 4 &$>$93\%& $<$2min  \\
                                       & 4 $\times$ 4-\textbf{XOR APUF}\tnote{\textdagger} & 4 &$>$93\%& $<$10min  \\
                                       & 5 $\times$ 3-\textbf{XOR APUF}\tnote{\textdagger} & 4&$>$94\%& $<$20min  \\
                                       & 2 $\times$ 3-\textbf{XOR APUF}\tnote{\textdagger}, 2 $\times$ 4-\textbf{XOR APUF}\tnote{\textdagger},1 $\times$ 5-\textbf{XOR APUF}\tnote{\textdagger}& 5 &$>$92\%& $<$10min  \\
\cmidrule{2-5}
\multirow{2}{*}{\uppercase\expandafter{\romannumeral 2}}& 4 $\times$ (2-1)-\textbf{Homo. XOR FF-APUF}\tnote{\textdagger\textdagger} & 4 &$>$95\%& $<$20min  \\
                                       & 4 $\times$ (1-3)-\textbf{Homo. XOR FF-APUF}\tnote{\textdagger\textdagger} & 4 &$>$92\%& $<$30min  \\
\cmidrule{2-5}
\multirow{2}{*}{\uppercase\expandafter{\romannumeral 3}}& 4$\times$ (3-1)-\textbf{ Hete. XOR FF-APUF}\tnote{\textdagger\textdagger} & 4 &$>$92\%& $<$30min  \\
                                       & 4$\times$ (2-2)-\textbf{ Hete. XOR FF-APUF}\tnote{\textdagger\textdagger} & 4 &$>$95\%& $<$20min  \\
\cmidrule{2-5}
\multirow{4}{*}{\uppercase\expandafter{\romannumeral 4}}& 4 $\times$ (1,5)-\textbf{Interpose PUF}\tnote{\textdaggerdbl} & 4 &$>$96\%& $<$30min  \\
                                       & 2 $\times$ 3-\textbf{XOR APUF}\tnote{\textdagger}, 2 $\times$ 4-\textbf{XOR APUF}\tnote{\textdagger},1 $\times$ (1,5)-\textbf{Interpose PUF}\tnote{\textdaggerdbl} & 5 &$>$96\%& $<$30min  \\
                                       &1 $\times$ (2-1)-\textbf{Hete. XOR FF-APUF},1 $\times$ (2-2)-\textbf{Hete. XOR FF-APUF},1 $\times$ (3-1)-\textbf{ Hete. XOR FF-APUF}\tnote{\textdagger\textdagger},1 $\times$ (1,5)-\textbf{Interpose PUF}\tnote{\textdaggerdbl}& 4 &$>$90\%& $<$30min  \\
                                       &1 $\times$ (1,5)-\textbf{Interpose PUF}\tnote{\textdaggerdbl},1 $\times$ (2,2)-\textbf{Interpose PUF}\tnote{\textdaggerdbl},1 $\times$ (3,3)-\textbf{Interpose PUF}\tnote{\textdaggerdbl}& 3 &$>$90\%& $<$30min  \\

\bottomrule
\end{tabular*}
\smallskip
\scriptsize
\begin{tablenotes}
\item[\textdagger] The parameter indicates the number of XOR stages;
\item[\textdagger\textdagger] The two parameters indicate the number of XOR stages and loops;
\item[\textdaggerdbl] The parameter indicates the number of parallel arbiter chains in up and lower layers.
\item[*] In the table, the amount of training data for each PUF is the same as the corresponding single task.
\end{tablenotes}
\end{threeparttable}
\end{table*}

\section{Generic Framework of Multiple PUFs}
\label{sec:mul}
In Section \ref{sec:single}, we have shown the generality of the proposed model that does not need to design specific structures for different types of PUFs, and no information apart from CRPs is needed. Then we evaluate another generic property of our proposed model on multiple tasks. To the best of our knowledge, multiple-task learning for multiple PUFs has not been studied before. In this section, we first show how multiple tasks learning challenges the existing modelling methods and compare them with ours. Then, we present our modelling accuracy results on different combinations of various types of PUFs.

\subsection{Setup}
In Section. \ref{sec:method}, we show the structure of a generic framework for Multiple-task learning and the strategy of choosing experts. In the experiments, we first set up several randomly chosen PUFs, then generated the challenges and used them to query all the PUFs and collect the responses to store in a list. We feed the challenges and the response list to the model. The model does not know from which PUF the data is from but creates one gate function, one tower layer and one output layer per group of responses. The model is expected to predict all the responses to unseen challenges for all the PUF inputs.

\subsection{Modelling Results Comparison and Summary}
We employed extra output layers on contrasting schemes listed above to enable their multi-task learning capability. After the modifications, their models look like the share-bottoms \cite{caruana1997multitask}, which is the most common way of enabling multiple-task learning. \textit{For \cite{Ruhrmair_modelling}, we cannot add output layers for a mathematical-model-based structure since they used multiplying layers and output directly the result after the activation function.} Due to the new methodology proposed in \cite{mishra2024calypso}, its publication shortly before the time of writing, and some issues with the released codebase (also reported in \cite{li2023could}), replication of it was not feasible within the scope of this work. We have instead referenced the provided results in \cite{mishra2024calypso} for the purposes of comparison, though this is limited by what tests were performed in the work referenced. For \cite{Mursi,aseeri2018machine}, we add one extra output layer. The results for modelling multiple XOR PUFs are shown in Table \ref{tab:results_multiple_comb}. The results show that simply applying multiple-task learning features on a proven feasible model can not balance between different tasks and will always fail one random task, which is described as the result of ``Negative Transfer" when the two tasks have low similarity or the experts cannot understand the correlation. In this case, multiple-task learning does not work and even the performance for training the single task is also degraded by the other conflicting task. However, as shown in Table \ref{tab:results_multiple_comb}, our proposed model can deal with different tasks with good performance compared to the same single task. For two 7-XOR APUFs, we need the same amount of CRPs for both and achieve an average accuracy of 98.04\% in 20 minutes. On average, we can save half of the training time, which will be efficient for the adversary modelling multiple PUFs. We also list more combinations of different numbers of various PUFs in Table \ref{tab:combinations} to show the flexibility of the proposed model. Specifically, we are testing the modelling ability for multiple PUFs \textit{where the PUFs are not all of the same type} e.g., some devices being attacked have FF-APUFs and others have Interpose PUFs, devices use XOR-APUFs of varying size, etc. We test several such combinations to demonstrate the consistency of the method but not every possible permutation to keep this experiment's complexity manageable. In particular, in comb. \uppercase\expandafter{\romannumeral 1}, we show the results of modelling four XOR APUFs with the same stages and different stages. In comb. \uppercase\expandafter{\romannumeral 2}, we show the results of modelling four homogeneous XOR FF-APUFs with different types. In comb. \uppercase\expandafter{\romannumeral 3}, we show the results of modelling four heterogeneous XOR FF-APUFs.  In comb. \uppercase\expandafter{\romannumeral 4}, we show the results for different combinations of XOR APUFs, (two kinds) XOR FF-APUFs, and Interpose PUFs. Compared to the results in Table \ref{tab:results_single}, $3\%$ of accuracy is traded off on average, and the maximum is around $8\%$. This shows that even for systems containing mixtures of completely different PUF designs, this approach consistently achieves viable attack capability (at least, for cases where all are delay-based PUFs).



\section{Conclusion}
In this study, we proposed a generic framework for modelling different delay-based PUFs. In this regard, we introduced a new notion called the Mixture-of-PUF-Experts (MoPE) Layer that enables attacks with minimal knowledge using the gate function and experts of PUFs. A realistic threat model has been considered where the Minimum Viable Adversary can only sniff the data in the network without knowing any other information about the communication objects. We showed successful attack results on XOR-APUFs, both homogeneous and heterogeneous XOR FF-PUFs and Interpose PUFs, without changing any settings of the model which answers \textbf{RQ 1}. Besides, we have also proposed an extended version of MoPE i.e., Multi-Gate Mixture-of-PUF-Experts. It enables multiple-task modelling on PUFs, which can capture the relationship between similar PUFs and accelerate the modelling process and answer \textbf{RQ 3}. We are the first to enable multiple-PUF attack capability of adversaries without sacrificing unacceptable loss of accuracy. To facilitate a fair comparison with the latest advancements in PUF modelling, we have undertaken analogous experiments to those conducted in previous studies such as \cite{Ruhrmair_modelling, Mursi, aseeri2018machine}. Experiments on different datasets, including simulated, biased silicon and unbiased silicon data, were performed to validate our methods. Here, we argue that our proposed models successfully solve \textbf{RQ 2} and will be helpful for the PUF community, especially when one comes up with a new PUF design and is willing to test whether their PUF is ML-MA secure without disclosing details on the PUF. In this research, we mainly considered the delay-based PUF(s); however, in the future, we would like to consider some other categories of PUFs.


\bibliographystyle{IEEEtran}
\bibliography{reference}


\end{document}